\documentclass[useAMS,usenatbib]{mn2e}
\usepackage{graphicx}
\title[Optical polarization angle and VLBI jet direction in the binary black
hole model of OJ287]{Optical polarization angle and VLBI jet
  direction in the binary black hole model of OJ287}
\author[Mauri J. Valtonen et al.]{Mauri J. Valtonen$^1$\thanks{E-mail:
    mvaltonen2001@yahoo.com}, Carolin
  Villforth$^{2,3}$ and Kaj Wiik$^2$ \\
  $^1$Helsinki Institute of Physics, FIN-00014 University of Helsinki, Finland\\
  $^2$Tuorla Observatory, Department of Physics and Astronomy, University of
  Turku, V\"ais\"al\"antie 20, FI-21500 Piikki\"o, Finland\\
  $^3$Space Telescope Science Institute, 3700 San Martin Drive, 21218 Baltimore,
  Maryland, USA\\}
\begin{document}

\date{Accepted. Received; in original form }

\pagerange{\pageref{firstpage}--\pageref{lastpage}} \pubyear{2011}

\maketitle

\label{firstpage}

\begin{abstract}
  We study the variation of the optical polarization angle in the blazar OJ287
  and compare it with the precessing binary black hole model with a 'live'
  accretion disk. First, a model of the variation of the jet direction is
  calculated, and the main parameters of the model are fixed by the long term
  optical brightness evolution. Then this model is compared with the
  variation of the parsec scale radio jet position angle in the sky. Finally,
  the variation of the polarization angle is calculated using the same model,
  but using a magnetic field configuration which is at a constant angle relative
  to the optical jet. It is found that the model fits the data reasonably well
  if the field is almost parallel to the jet axis. This may imply a
  steady magnetic field geometry, such as a large-scale helical field.
\end{abstract}

\begin{keywords}
galaxies: active -- galaxies: jets -- BL Lacertae objects: individual: OJ287.
\end{keywords}

\section{Introduction}
The blazar OJ287 shows an interesting light curve where optical outbursts
follow each other in roughly a 12 yr sequence (Sillanp\"a\"a et al. 1988). The
light curve has two significant periodicities, the 12 yr period as well as a
60 yr period (Valtonen et al. 2006a). The simplest explanation of such a
doubly periodic system is that it is a binary black hole (BBH) system
(Katz 1997) where the accretion disk of the primary is perturbed by a
companion on a 12 yr orbit, while the larger period represents a precession
cycle. Sundelius et al. (1997) presented a detailed study of this case, and
noted that the disk is tidally perturbed during the close encounters. It is
reasonable to assume that the tidal perturbations affect the accretion flow,
and thus create a 12 yr pattern of increased brightness. The model of
Sundelius et al. (1997) is illustrated in Figure~1. The massive central black hole lies in the centre of the accretion disc. The jet emanating from the black hole (not drawn in the figure) is taken to be perpendicular to the disc.

In the BBH model the primary is a massive black hole belonging to the
upper end of the black hole mass function (Ghisellini et al. 2009, Sijacki et
al. 2009, Kelly et al. 2010, Trakhtenbrot et al. 2011). The host galaxy of
OJ287 is bright, $M_{K}\sim-28.9$ (Wright et al. 1998). The primary mass of
$1.8\times10^{10}$ solar mass places OJ287 almost exactly on the black hole
mass - host galaxy magnitude relation extended from smaller galaxies and black
holes (Kormendy \& Bender 2011). The secondary moves in an eccentric orbit of
eccentricity $e\sim0.7$, and it impacts the accretion disc of the primary
twice during its 12 yr orbit. The impact points vary from orbit to orbit, and
by studying this variation one may determine the rate of the relativistic
precession and other parameters of the orbit (Valtonen 2007). The flares
arising from disc impacts are easily recognised by the sudden rise of the
optical flux and by their generally short duration in comparison with tidal
outbursts (Valtonen et al. 2008a, 2009). The mass of the secondary black hole
is $\sim10^8$ solar mass, small enough that the accretion disc remains stable
in spite of the repeated impacts. The accretion disc is modeled as a magnetic
$\alpha_g$ disc of Sakimoto \& Coroniti (1981). The binary orbit is taken to
be nearly perpendicular to the plane of the disc; therefore the jet (not
illustrated in Figure~1) lies close to the binary plane.

The future optical light curve of OJ287 was predicted from 1996 to 2030;
during the first fifteen years OJ287 has followed the prediction with amazing
accuracy, producing five outbursts at expected times, of expected light curve
profile and size (Valtonen et al. 2011). In this theory the orbit solution of
Lehto $\&$ Valtonen (1996) was used, but because Sundelius et al. (1997) use a
'live' disc model, the disk is lifted up by the approaching secondary (Ivanov
et al. 1998), and the outbursts related to impacts happen earlier than
expected in a 'rigid' disc model. For example, in the 'live' disc model the
2005 disk impact is about 6 months ahead of the impact in the 'rigid' disc
model, and thus the 2006 April outburst is shifted to 2005 October, as was
subsequently observed (Valtonen et al. 2006b,2008b).

In this paper we calculate another property of a 'live' accretion disc, the
change of its orientation as a function of the phase of the binary
orbit. Since the changes in the disc orientation are small, Sundelius et
al. (1997) did not make an effort to calculate them. The model calculations
are then compared with the historical VLBI data of the parsec scale jet
orientation as well as with the optical polarization data collected by
Villforth et al. (2010), as it is possible that the changing disc orientation
is reflected in the orientation of the jet. The jet orientation, in turn, may
affect the direction of polarization in the optical emission of the jet.

\begin{figure}
 \includegraphics[width=\columnwidth]{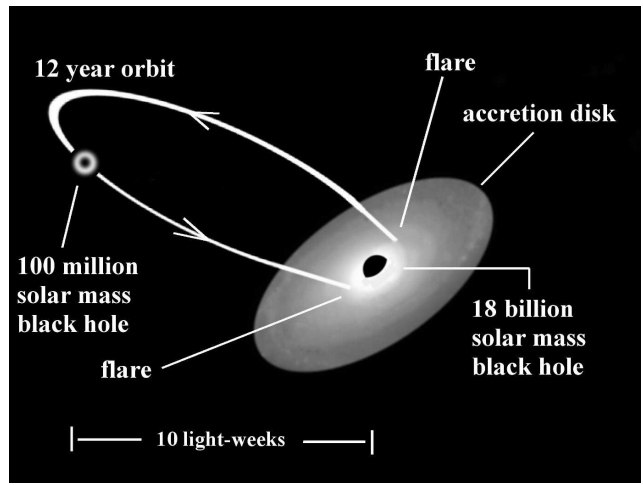} 
 \caption{An illustration of the Sundelius et al. (1997) model. The jet is not shown, but it is taken to lie along the rotation axis of the accretion disc.}
 \label{fig1}
\end{figure}

\section{Optical polarization}
In this paper we concentrate on the variation of the position angle of
polarization in OJ287, as this is a quantity which may be modeled in the
Sundelius et al. (1997) framework. The data have been published and are
illustrated in Figure 14 of Villforth et al. (2010). They include data points
both from the authors' own campaign from 2005 to 2009, and earlier
measurements. From the point of view of our theory, we are interested in the
long term behaviour of the mean polarization angle. Therefore we take averages
per each calendar year. They are listed in column 2 of Table 1, together with
the standard deviation of the scatter per calendar year in column 3. The last
two columns give the number of observations used and the standard error of the
mean (Columns 4 and 5, respectively).

\begin{table}
 \centering
  \caption{Mean optical polarization angle and its uncertainty.}
  \begin{tabular}{@{}llrrrrrrrrrr@{}}
  \hline
   Year&Circ PA&STD& N&STD/sqrt(N)\\
   \hline
   1971 & 12 & 22 & 11 & 7\\
   1972 & 87 & 23 & 95 & 2\\
   1973 & 100 & 10 & 33 & 2\\
   1974 & 75 & 7 & 21 & 2\\
   1975 & 85 & 9 & 36 & 1\\
   1976 & 81 & 9 & 20 & 2\\
   1977 & 100 & 10 & 21 & 2\\
   1978 & 78 & 4 & 9 & 1\\
   1979 & 88 & 8 & 23 & 2\\
   1980 & 78 & 3 & 6 & 1\\
   1981 & 134 & 12 & 11 & 4\\
   1982 & 62 & 14 & 35 & 2\\
   1983 & 101 & 17 & 218 & 1\\
   1984 & 145 & 26 & 101 & 3\\
   1985 & 160 & 17 & 14 & 5\\
   1986 & 29 & 26 & 18 & 6\\
   1987 & 100 & 11 & 13 & 3\\
   1988 & 65 & 20 & 33 & 3\\
   1989 & 127 & 7 & 64 & 1\\
   1990 & 114 & 8 & 174 & 1\\
   1991 & 93 & 13 & 129 & 1\\
   1992 & 106 & 3 & 48 & 1\\
   1993 & 62 & 6 & 29 & 1\\
   1994 & 137 & 25 & 105 & 2\\
   1995 & 157 & 22 & 75 & 3\\
   1996 & 176 & 18 & 155 & 1\\
   1997 & 169 & 8 & 95 & 1\\
   2005 & 174 & 15 & 15 & 4\\
   2006 & 171 & 16 & 120 & 1\\
   2007 & 164 & 16 & 129 & 1\\
   2008 & 166 & 21 & 104 & 2\\
   2009 & 170 & 16 & 31 & 3\\
   \hline
 \end{tabular}
\end{table}

A few specific points should be noted about the calculation of the
average. First, calculating means and standard deviations for directional data
can be challenging, especially if the scatter is large. To avoid problems, we
use directional statistics for all directional data (Mardia 1975). This is
implemented by use of the SciPy module scipy.stats.morestats
(www.scipy.org). Second, the individual measurement errors are generally much
smaller than the rms scatter. Thus the scatter reflects a genuine variation of
the polarization angle, but since our model does not handle short time scale
variations, we will not try to model them. Figure~2 shows the
resulting mean values as squares and the circular standard deviation as error
bars.

\begin{figure}
\includegraphics [angle=270, width=\columnwidth]{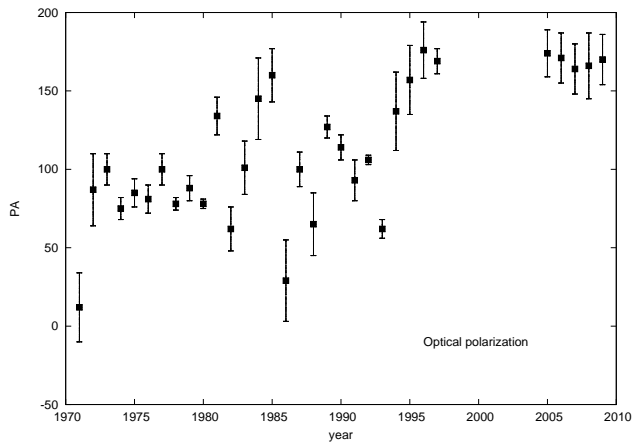} 
\caption{The evolution of optical polarization angle in OJ287 together with
  the circular standard deviation (error bars). Annual mean values are shown.}
\label{fig2}
\end{figure}

\section[]{Dynamical model}
Our model is the same as in Sundelius et al. (1997) except that the error bars
in the orbit model have been recently narrowed down after new outburst timings
have been included in the solution (Valtonen et al. 2010a). However, for the
purposes of this paper the increased accuracy is of no consequence since the
Sundelius et al. (1997) model was already quite accurate. Similarly to
Sundelius et al. (1997), the disc of the primary is modeled by non-interacting
particles. This limitation is not as bad as it may seem; Sundelius et
al. began their simulations with a self-interacting disc, as reported e.g. in
Sillanp\"a\"a et al. (1988), but they soon found out that for the kind of
binary orbit considered here, where the disc plane and the orbit plane are
perpendicular to each other, the inclusion of self-interaction only increases
calculation time greatly without producing significantly different results.

In the present simulation the number of disc particles is 37200. They are
placed in circular orbits around the primary, with orbital radii ranging from
8 Schwarzschild radii to 20 Schwarzschild radii of the primary. Varying the
inner and outer radii of the disc was not found to influence the result
significantly. For every particle, and for every time step, we calculate the
orbital elements of the orbits with respect to the primary. The elements are
averaged per calendar year. As there are many integration steps per year,
typically each annual mean is based on the average of between 2 million and 10
million values. By varying the number of particles it was found that the mean
values generated in this way are very robust.

We are interested primarily in the mean orientation of the disc, in the region
affected by the secondary. Thus only two orbital elements are of consequence,
the inclination of the mean disc $i$ and the corresponding ascending node
$\Omega$. The fundamental plane is taken to be the plane of the binary (x-y
-plane) at the beginning of the calculation, in year 1856. In the latest orbit
model which includes the spin of the primary black hole this plane evolves
slowly (Valtonen et al. 2010a). However, the evolution is so slow in the time
scale that we are considering that it may be viewed as an invariable
plane. The disc is initialized such that the inclinations are typically around
$\sim90^{\circ}$, i.e. particles move in the x-z -plane. For the convenience
of discussion, we show in the following figures the quantity $i-\frac{\pi}{2}$
instead of the inclination $i$. The ascending node is measured from a
fundamental line (x-axis) in the plane of the binary. The disc is initially
loaded such that $\Omega=\pi$, i.e. particles cross the x-y -plane from the
underside (negative z-axis) to the upperside (positive z-axis) at the negative
x-axis. Figure~3 shows the evolution of these two orbital
elements. Here and in the following, $\Omega$ is replaced by $\Omega-\pi$. In
practice, the variation of the inclination is very small in comparison with
$\Omega$. Why this is so will become clear below (Eqs. 1 \& 2).

At this point a free parameter enters our calculation. The sound crossing time
of the accretion disc (the region we are considering) is about 10 years. Thus
it takes about 10 years to generate physically significant mean values; we do
this by taking a sliding average of the surrounding (future and past) values
at each annual point. The number of years used in the average is left as a
free parameter. In practice we have noticed that taking a sliding mean between
7 and 11 years gives us enough information to see how this free parameter
influences our results (see Figure~3).

\begin{figure}
\includegraphics [angle=270, width=\columnwidth]{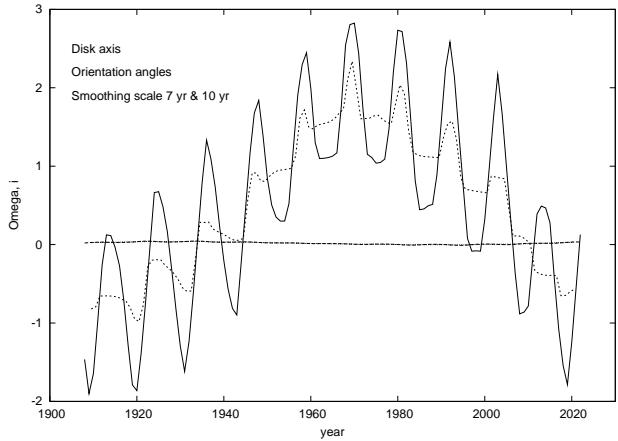}
\caption{The evolution of the mean disk in the binary model. The lines with
  the larger amplitude refer to the ascending node $\Omega-\pi$, with a 7 yr
  (solid line) and a 10 yr (dashed line) average. The almost constant lines
  refer to the complement of the inclination $i-\frac{\pi}{2}$.}
\label{fig3}
\end{figure}

Figure~3 shows two cycles: a 120 yr precession cycle and the twelve
year 'nodding' cycle. With the sliding mean of 10 years, the 120 yr precession
cycle dominates, while going towards the 7 yr sliding average the 12 yr
nodding cycle becomes more pronounced. Katz (1997) explained these two
frequencies in a physical model. In analogy to Hercules X-1 and SS 433, the
precession was attributed to the torque exerted by a companion mass on an
accretion disc. Our model is exactly the same, except Katz (1997) chose 12
years and 1.2 years as the periods of the two cycles. However, as mentioned
above, in the power spectrum analysis of the light curve of OJ287 the dominant
peaks occur at 60 yr and 12 yr (Valtonen et al. 2006a). The observed 60 yr
cycle may be only half of the full cycle since Doppler boosting is increased
twice during each 120 yr cycle if the viewing angle is small. The model of
Katz (1997) is therefore easily applied to the present system. For the
increase of cyclic periods by a factor of ten over Katz (1997), and keeping
the gravitational radiation lifetime the same, the masses have to be increased
by more than an order of magnitude from the $10^{9}$ solar mass scaling by
Katz (1997). This fits nicely with the mass scale used in our model where the
primary mass is $\sim 1.8\times10^{10}$ solar mass and the secondary $\sim
1.4\times10^8$ solar mass.

These cycles are also well known in three-body orbit dynamics. In the twice
orbit-averaged hierarchical three-body problem the first order equations for
the evolution of inclination $i$ and the ascending node $\Omega$ are written
\begin{eqnarray}
\frac{\mathrm{d}i}{\mathrm{d}\tau}=0
\end{eqnarray}
\begin{eqnarray}
\frac{\mathrm{d}\Omega}{\mathrm{d}\tau}=-\frac{3}{4} \cos i
\end{eqnarray}
(Valtonen \& Karttunen 2006) where $\tau$ is the normalized time coordinate,
\begin{eqnarray}
\tau\sim\frac{P}{P_e}\frac{t}{P_e}
\end{eqnarray}
and $t$ is time, and $P$ and $P_{e}$ are the inner and outer periods,
respectively. The actual motions are cyclical: the cycle is called the Kozai
cycle and its period is $P_{Kozai}\sim 2 P_{e}^2/P$. For our case $P_{e}\sim
12$ yr, $P\sim P_{e}/5$, and $P_{Kozai}\sim 120$ yr.  For inclinations close
to $90^{\circ}$, $\cos i \sim \frac{\pi}{2}-i$. The amplitude of the $\Omega$
cycle may be estimated by putting $\mathrm{d}\tau\sim 1 $ in Eq. 2; thus the
maximal $\mathrm{d}\Omega \sim \frac{\pi}{2}-i$, i.e. it is of the order of one
degree in models where the inclination is also within one degree of
$90^{\circ}$. Note that the averaging which is done numerically in our
simulations is analogous to the analytical orbit averaging in the dynamical
theory.

\section{The jet}

\subsection{Modeling the orientation of the optical jet}

Presumably the optical emission of OJ287 arises at most times in its
jet. Therefore we need to make further assumptions about the disk/jet
connection. The disk/jet connection is very much an open problem, and
therefore we make a simple assumption: let the jet lie exactly along the
rotation axis of the disk and let any bending of the jet be negligible up
  to parsec scale, i.e. the effect of jet instabilities is omitted here.
With this assumption we generate a time sequence of jet orientations
from Figure~3. The jet orientation shows up in observations in
several ways. First, Doppler boosting generates the base level of optical
brightness for OJ287. The 60 yr cycle mentioned earlier may be related to this
variation. Thus we take observations from the historical light curve (see
e.g. Valtonen et al. 2010b) and remove the well known outburst peaks. The
resulting base level light curve is shown in Figure~4. In generating
this light curve the highest optical V-magnitude was selected from each 0.1 yr
interval of time.

\begin{figure}
\includegraphics [angle = 270, width=\columnwidth]{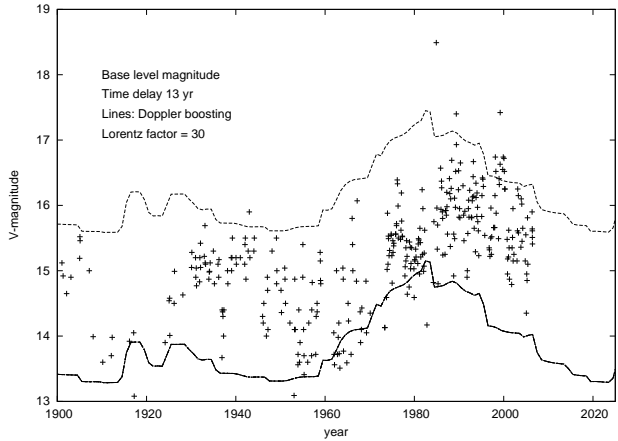}
\caption{The variation of the base level brightness in OJ287. The V-magnitudes
  are the minimum measured values per 0.1 yr time interval. Curves represent
  the effects of Doppler boosting on the base level magnitude.}
\label{fig4}
\end{figure}

Figure~4 has also two lines representing the effect of the Doppler
boosting on the optical brightness, vertically displaced by about two
magnitudes, which are generated using the data in Figure~3. In order
to carry out the transformation from Figure~3 to Figure~4 it
is necessary to adopt four free parameters: two parameters of the viewing
angle of the observer, the Lorentz factor $\Gamma$ in the jet, and the time
delay between the instantaneous change of the disk plane and the transmission
of this information to the central axis and the jet. The values for the
parameters are obtained by fitting the curves to the range of data points in
Figure~4.

In order to calculate the viewing angle, we parameterize the plane
perpendicular to the line of sight by the same two angles, inclination $i_{0}$
and ascending node $\Omega_{0}$, as those used to describe the disk
plane. Thus the direction to the observer is represented by two horizontal
lines in Figure~3. For any moment of time we measure the vertical
distance between the line $\Omega_{0}-\pi$ and the $\Omega-\pi$ curve as well
as the distance of the $i_{0}-\frac{\pi}{2}$ line and the $i-\frac{\pi}{2}$
curve. This gives us the two components of the viewing angle as a function of
time. The $i-\frac{\pi}{2}$ component of the viewing angle is practically
constant. In this way we generate sets of viewing angles as a function of
time, and shift them forward by some value $\Delta t$ which should be of the
order of the sound crossing time in the disk. For each moment of time we may
then calculate the Doppler boosting factor, assuming some value of $\Gamma$.

In order to produce a 60 yr periodic component out of the basic 120 yr cycle,
we need to choose the viewing angle so that $\Omega_{0}$ is inside the
variability range of $\Omega$ in Figure~3. The curves in
Figure~4 are based on $\Omega_{0}-\pi=0^{\circ}.4 $,
$i_{0}-\frac{\pi}{2}=-0^{\circ}.4$, $\Delta t = 13$ yr and $\Gamma=30$. This
fit does not take into account other factors besides the steady jet
contributing to the optical brightness. Thus the value of $\Gamma$ is only
indicative of the general magnitude of the Lorentz factor; it is not
significantly different from the values obtained by other means ($\Gamma\sim
20$, Jorstad et al. 2005). Our range of viewing angles is between $1^{\circ}$
and $2^{\circ}$, again in agreement with findings in other studies
(L\"ahteenm\"aki and Valtaoja 1999, Jorstad et al. 2005, Savolainen et
al. 2010).

\subsection{Orientation of the VLBI jet}

\begin{table}
 \centering
 \caption{Historical data showing the mean direction of OJ\,287's VLBI jet
   within the first 1\,mas from the core. The full table is available as
   Supporting Information with the online version of this article.}
 \label{jetpa}
  \begin{tabular}{@{}cccccc@{}}
  \hline
   Obs. Date &  Array$^a$ & $\lambda$  &  PA   & $\Delta$PA$^b$ & Ref.$^c$\\
      (yr)   &            &   (cm)     & (deg) &  (deg)         &   \\ 
  \hline
  1981.95    & USVN       &     6      & -116  &  ...       & 1 \\
  1982.95    & USVN       &     6      & -113  &  ...       & 1 \\
  1985.51    & Geo        &     3.6    & -82   &   3        & 2 \\
  1985.75    & Geo        &     3.6    & -83   &   3        & 2 \\
  1985.89    & Geo        &     3.6    & -93   &   3        & 2 \\
  1986.27    & Geo        &     3.6    & -103  &   5        & 2 \\
  1986.50    & Global     &     6      & -100  &  ...       & 3 \\

  \hline
  \end{tabular}

  \medskip \flushleft $^a$\,USVN\,=\,U.S. VLBI Network, Geo\,=\,Geodetic VLBI
  experiment, VLBA\,=\,Very Long Baseline Array, VLBA+\,=\,VLBA and geodetic 
  stations, VSOP\,=\,VLBI Space Observatory Programme\\
  $^b$\,If no error estimate is available, then $\Delta\mathrm{PA} = 10^\circ$ 
  is assumed.\\
  $^c$\,References: (1) \citet{rob87}, (2) \citet{vic96}, (3) \citet{gab89},
  (4) \citet{gab96}, (5) \citet{tat99}, (6) \citet{fey96}, (7) VLBA 2cm Survey
  and MOJAVE program; \citet{kel98}, \citet{lis09a,lis09b}, (8) \citet{fey97}, 
  (9) \citet{jor01}, (10) \citet{ojh04}, (11) \citet{pin07},
  (12) \citet{gab01}, (13) \citet{tat04} \\

\end{table}

The second way to measure the jet rotation is to look at its position angle in
the sky. Unfortunately it is impossible to resolve the optical jet, but at
radio wavelengths the jet is observable in parsec scale with the VLBI and its
orientation is known since the early 1980s.

We have constructed the observed history of variations in the position
angle of OJ287's radio jet by collecting the available VLBI data from the
literature. The jet PA was defined as the mean PA within the first one
milliarcsecond (mas) from the VLBI core, which is the bright feature at one
end of the jet. The mean PA was measured by fitting an
  ordinary-least-squares bisector regression line to the jet component
positions if more than one such component\footnote{The word ``component''
  refers to 2-dimensional Gaussian flux profiles that are fitted to the VLBI
  data in order to parameterize the observed brightness distribution of the
  jet. The parameters of these models are typically reported alongside with
  the VLBI images in the literature.}  was present within 1\,mas from the
core. The regression line was forced to go through the core position and the
each component was assigned a positional uncertainty equivalent to 10\% of the
component's size convolved with the beam size (see Lister et al. 2009a). In the
least-squares fit the data points were weighted by the inverse square of their
positional uncertainty times the component's (normalized) flux density. If
only one component was present within 1\,mas from the core, the PA of this
knot was assumed as the jet PA. Table~2 lists 140 individual jet PA
measurements obtained in this manner and covering a time period from 1981 to
2009. These observations were made at 2, 3.6, and 6\,cm wavelengths. We chose
these wavelengths because within this range the jet PA seems to be very weakly
wavelength-dependent, and because there are abundant historical data available
through the geodetic observations (3.6\,cm) and the VLBA 2\,cm / MOJAVE
Surveys (Kellermann et al. 1998,Lister et al. 2009a). We did not try to back-extrapolate the ejection
epochs of the knots, since it easily leads to ambiguous results for
inhomogeneous data sets with time gaps. The proper motions close to the core
in OJ287 are $\sim0.5-0.8$\,mas\,yr$^{-1}$ meaning that we measure the changes
in the PA with a delay of $\sim 1 - 2$\,yr (Lister et al. 2009a).

Figure~5 presents the long-term evolution of the VLBI jet's position
angle. The values there are weighted annual means of $2 - 6$\,cm
measurements. The error bars represent either the standard deviation of the
scatter in a given year (multiple observations per year) or the uncertainty of
a single measurement. As mentioned earlier, stacking 2, 3.6, and 6\,cm data
together does not add a bias to the PA curve, since the average difference in
(quasi-)simultaneously measured PAs between these wavelengths is close to
zero. 

The projected jet PA presented in Figure~5 varies with a total
amplitude of $\sim50$ degrees, it has an overall decreasing trend, and it
shows two significant ``spikes'' separated by about a decade. Tateyama \& Kingham 2004
have earlier reported a varying jet PA in OJ287 and proposed that this may be
due to the precession of a ballistic jet. Comparison of Figure~5 to
Figure~6 in \citet{tat04} shows, however, that this specific periodic model
does not fit the observed variations in jet PA when data spanning three
decades is considered. Moor et al. (2011) show another recent compilation of the PA data with a 12 yr cyclic structure together with the overall long term trend.

\begin{figure}
\includegraphics [angle = 270, width=\columnwidth]{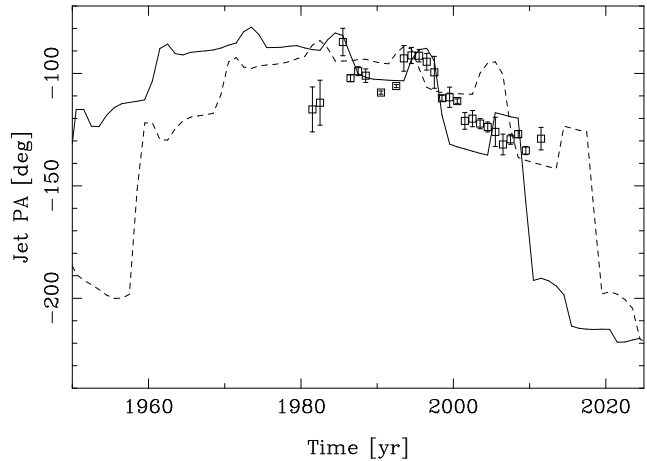}
\caption{The variation of the position angle of the parsec scale radio jet in
  the sky. Squares are annual averages, and the error bars represent
  uncertainties. The dashed line refers to the model with a 10 yr average and $\Delta t = 13$ yr. If $\Delta t = 4$ yr, as discussed in the text, then the model line is moved backward, and after a slight adjustment of the vertical scale, is shown as a solid line.}
\label{fig5}
\end{figure}

The lines in Figure~5 are the projected position angles in the sky
calculated from our model. They have been drawn for the same parameter values as
the lines in Figure~4, except that for the solid line the time delay $\Delta t$ is
only 4 years. The free parameters here are the time delay and the zero point
of the projected jet orientation. The 13 yr time delay model and the PA measurements agree
in general outline, with the exception of the two earliest data points, which are
from four-station experiments with the U.S. VLBI array. The poor $(u,v)$
coverage of these observations leaves the jet orientation more uncertain than
in the rest of the data.

The model predicts that the jet PA should rapidly turn clockwise by $\sim50$
degrees within the next ten years. Also, the model is coming to a minimum in the
angle between the jet and our line-of-sight. As the viewing angle approaches
the half-opening angle of the jet, the appearance of the VLBI jet can change
significantly: the apparent opening angle increases, any bending is
exacerbated, and differential Doppler boosting may affect the observed
structure. Based on a sequence of 7\,mm VLBA images Agudo et al. (2010,2011)
recently reported a greatly changed PA in the innermost part of OJ287's jet
during 2008--2010. They gave a new jet PA of $\sim -200^{\circ}$ or $\sim
-20^{\circ}$, depending on the identification of the core
component.  Agudo et al. (2010) propose that the jet is indeed crossing our
line-of-sight and the sudden change in the PA may be due to a moving kink in
the jet with an almost zero viewing angle. One cannot, however, directly
compare the PA values reported by them with those presented in
Figure~5, since the PAs given in  Agudo et al. (2010,2011) refer to the
mutual orientation of the two innermost bright features at 7\,mm that are
separated by only $\sim0.2$\,mas, whereas the PAs shown in Figure~5
refer to the mean orientation of the jet within the first 1\,mas in the
2-6\,cm images. Therefore, further VLBI observations are needed to test if the
predicted change in the jet orientation takes place.

If the time delay turns out to be shorter than in the optical variations, it may imply that the
information on the change of jet orientation is transmitted to the axis via
the hot corona surrounding the accretion disk, with a speed of sound
(actually, Alfven speed) which is much greater than the sound speed in the
disk. The value of $\Gamma$ does not enter this calculation.

\section{Optical polarization angle}
In the following we make the simplifying assumption that the symmetry axis of the magnetic field structure in the jet is at a constant angle relative to
the jet axis. This basically requires either a reasonably stable
vector-ordered magnetic field in the jet or a long-lived stationary shock in
the region where the optical emission originates. Thus when the jet points in
different directions, so does the electric field vector of the oscillating electrons. Consider first the case where the electric field vector is almost parallel to the jet. The
electric field vector has its own viewing angle which is in general different
from the viewing angle of the jet. Let the plane which is perpendicular to the
electric field vector be specified by angles $\Omega_1$ and $i_1$. We will now
consider determining these parameters when $\Delta t = 13$ yr in the optical
part of the jet.

Figure~6 shows the polarization angle data together with a fit to a
model with $ \Omega_1-\pi=1^{\circ}.6$ and
$i_{1}-\frac{\pi}{2}=0^{\circ}.25$. This means that the jet and the electric
field vector are at an angle of $1^{\circ}.4$ relative to each other. An
  almost parallel EVPA with respect to the jet axis could be generated by a
large-scale helical magnetic field with a dominant toroidal component or a
compressed magnetic field of a long-lived (oblique) standing shock front in
the jet (see Lyutikov et al. 2005 for a discussion of jet polarization).

However, there is a $90^{\circ}$ shift in the PA given by this model with respect to the best fit with observations. This leads us to another possibility, namely that the poloidal magnetic field component dominates in the section of the jet where the optical emission comes from. In that case the PA that we are plotting is the direction of the magnetic field, with a small but significant deviation from the jet line. Then we should add $90^{\circ}$ to the PA values in our model in order to compare with the observed polarization angle (Laing 1981, Lyutikov et al. 2005). It is not possible to decide $\it{apriori}$ which case we should have in OJ287 since it depends sensitively on the physical structure of the jet. However, it is generally thought that the optical emission arises rather close to the central engine (e.g. Agudo et al. 2011). Since the poloidal magnetic field component increases faster than the toroidal field component towards the centre, it is likely that in our case the poloidal field should dominate, and we should add $90^{\circ}$ to the PA values, as we do in Figure 6.

  We notice that the model
describes well the general feature of the polarization angle evolution,
noticed by Villforth et al. (2010), that the position angle has changed in an
almost steplike fashion from $\sim -90^{\circ}$ to $\sim 0^{\circ}$ in the
early 1990's. The prediction of the model is that this position angle should
hold on during the coming decades. Note that the model assumes optically thin conditions in the jet and thus the $90^{\circ}$ jump in PA around 1995 has nothing to do with the changes in optical depth. Neither do we need a dramatic change of the magnetic field configuration in the jet at that time, as suggested by Villforth et al. (2010). Quite opposite, our model assumes a constancy of the jet, and looks only at the PA changes due to the changing viewing angle.  

In some years the theoretical line goes well outside the error range of the
annual average. We should note that especially in mid 1980's the PA varied
over the whole angular range even in one year (see Figure 14 of Villforth et
al 2010), and the calculation of the average value is very sensitive to the
$\pm 180^{\circ}$ ambiguity in the observed polarization angles. For example, by a different choice of the quadrant for just one or two observations, the position angles $100^{\circ}$ in
1971, $116^{\circ}$ in 1984, $95^{\circ}$ in 1985, $90^{\circ}$ in 1986 and
$102^{\circ}$ in 1994 can also be found. In all of these cases the
correction goes to the direction of improving the fit with the theoretical
line. However, for the other years such uncertainty does not exits, and the
remaining difference with the theory must be considered real. We have also
broken down the data in semi-annual boxes, but it does not help to clarify the
situation with respect to the most uncertain points.

\begin{figure}
\includegraphics [angle=270, width=\columnwidth]{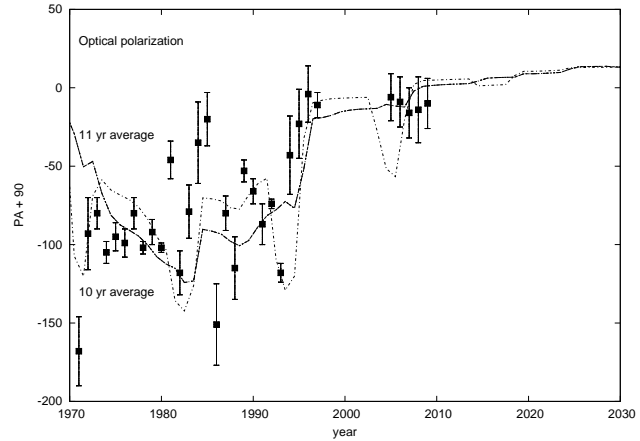}
\caption{The evolution of the optical polarization angle compared with the
  model. The curves are for a 10 yr (dash-dot line) and for an 11 yr (dashed
  line) average. Observations refer to the values in Table 1 minus
  $180^{\circ}$. The theoretical lines are shifted by adding $90^{\circ}$; there is a $90^{\circ}$ uncertainty in the polarization angle due to unknown physical conditions in the jet (Lyutikov et al. 2005).}
\label{fig6}
\end{figure}

\section{Conclusions} 

  Using a number of simplifying assumptions, we have calculated how the
  orientation of OJ287's relativistic jet is expected to vary in the binary
  black hole model of Sundelius et al. (1997). Even though the model does not
  include a detailed physical treatment of the jet/disk connection, it agrees
  with the observed radio jet orientation as well as optical polarization
  evolution of OJ287 in general outline. The scatter in the polarization
  position angle is large which makes a very specific comparison
  difficult.

The view which we have arrived at is that in many respects OJ287 can be
modeled as a binary system, a scaled up version of Her X-1 or SS433. In
  the presented model of OJ287, the jet sweeps across the line of sight in a
periodic manner, and this motion creates brightness variations by varying
Doppler boosting as well as shows up as a varying projection of the jet. While
the projected radio jet is actually observed, in optical we gain evidence of
the beam sweeping by the varying optical polarization angle. When the
analytical model of Katz (1997) is scaled up for the observed cycle
frequencies, it explains well the numerical simulation results of this
work. The presented model makes predictions about the future trends of
  the observables discussed in this paper and therefore it can be easily
  further tested by continuing the VLBI and polarization monitoring of OJ287.

\section*{Acknowledgments}
We would like to thank Tuomas Savolainen who has compiled the information in Table 2 and has generously allowed us to use it. This compilation has made use of data from the MOJAVE
database that is maintained by the MOJAVE team (Lister et al., 2009, AJ, 137,
3718). T.S. has also made helpful comments on the manuscript. We also acknowledge the comments of the referee which have improved the presentation.

\label{lastpage}

\end{document}